\documentclass[12pt]{article}
\usepackage{graphicx}

\newsavebox{\sboxpubnumber}
\newsavebox{\sboxpubdate}
\newcommand{\pubdate}[1]{\begin{lrbox}{\sboxpubdate}{#1}\end{lrbox}}

\newcommand{\Title}[1]{\begin{center} {\Large #1 } \end{center}}
\newcommand{\Author}[1]{\begin{center}{ \sc #1} \end{center}}
\newcommand{\Address}[1]{\begin{center}{ \it #1} \end{center}}

\newenvironment{Abstract}{\begin{quotation}  }{\end{quotation}}
\newenvironment{Presented}{\begin{quotation} \begin{center}
             PRESENTED AT\end{center}\bigskip
      \begin{center}\begin{large}}{\end{large}\end{center}
      \end{quotation}}
\newcommand{\Acknowledgements}{\bigskip  \bigskip \begin{center} \begin{large}
             \bf ACKNOWLEDGEMENTS \end{large}\end{center}}
\begin{document}

%%%%%%%%%%%%%%%%%%%%%%%%%%%%%%%%%%%%%%%%%%%%%%%%%%%%%%%%%%%%%%%%%%%%%%%%
%%
%% START EDITING HERE!
%%
%%%%%%%%%%%%%%%%%%%%%%%%%%%%%%%%%%%%%%%%%%%%%%%%%%%%%%%%%%%%%%%%%%%%%%%%
\begin{titlepage}
\pubdate{\today}                    %fill in the date
%\pubnumber{XXX-XXXXX \\ YYY-YYYYYY} %preprint number(s)

\vfill \Title{Probing the dark energy with redshift space quasar
clustering distortion}
\vfill \Author{M. O. Calv\~ao, J. R. T de Mello Neto and I. Waga}
\Address{Universidade Federal do Rio de Janeiro\\
         Instituto de F\'\i sica, C. P. 68528\\
         CEP 21945-970 Rio de Janeiro, RJ\\
         Brazil}
\vfill
%\andauth
%\vfill
%\Author{J. R. T. de Mello Neto and I. Waga}
%\Address{Department, Institute \\
%         Postal address}
%\vfill
\begin{Abstract}
We have run Monte Carlo simulations, for quasar clustering
redshift distortions in the Two-Degree Field QSO Redshift Survey
(2QZ), in order to elicit the power of redshift distortions
(geometric Alcock-Paczy\'nski and linear kinematic) to constrain
the cosmological density and equation of state parameters,
$\Omega_{m0}, \Omega_{x0}, w$, of a pressureless matter + dark
energy model. It turns out that, for the cosmological constant
case ($w=-1$), the test is especially sensitive to the difference
$\Delta:=\Omega_{m0}-\Omega_{\Lambda 0}$, whereas for the
spatially flat case ($k=0$), it is quite competitive with SNAP and
DEEP, besides being complimentary to them; furthermore, we find
that, whereas not knowing the actual value of the bias does not
compromise the correct recovering of $\Delta$, taking into account
the linear velocity effect is absolutely relevant, all within the
$2\sigma$ confidence level.
\end{Abstract}
\vfill
\begin{Presented}
    COSMO-01 \\
    Rovaniemi, Finland, \\
    August 29 -- September 4, 2001
\end{Presented}
\vfill
\end{titlepage}
\def\thefootnote{\fnsymbol{footnote}}
\setcounter{footnote}{0}

%%%%%%%%%%%%%%%%%%%%%%%%%%%%%%%%%%%%%%%%%%%%%%%%%%%%%%%%%%%%%%%%%%%%%%%%
% The document starts here
%%%%%%%%%%%%%%%%%%%%%%%%%%%%%%%%%%%%%%%%%%%%%%%%%%%%%%%%%%%%%%%%%%%%%%%%
\section{Introduction}

The new millennium has ushered in a golden era for cosmology,
driven by a flood of high quality observational data, from
supernovae \cite{Riess99,Perlmutter99,snap} to cosmic microwave
background \cite{deBernardis00,Balbi00,Pryke01,map,planck},
passing through galaxies and quasars \cite{2df,sdss}, to mention
just a few. All of them favor a spatially flat cosmological model,
with a nonrelativistic matter with density parameter $\Omega
_{m0}\simeq 1/3$ and a negative-pressure dark energy component
with density parameter $\Omega _{x0}\simeq 2/3$. The exact nature,
however, of this dark energy is not currently well understood,
possible alternatives being a vacuum energy or cosmological
constant ($\Lambda$) or a dynamical scalar field (quintessence)
\cite{Ratra88,Frieman95,Caldwell98,Ferreira98}. An important task
for present cosmology is thus to find new methods that can probe
the amount of dark energy present in the Universe as well as its
equation of state. These new methods may constrain distinct
regions of the parameter space and are usually subject to
different systematic errors.

The test we focus on here is the one suggested by Alcock and
Paczy\'nski (hereafter AP)\cite{Alcock79}, which has attracted a
lot of attention during the last years
\cite{Ryden95,Ballinger96,Matsubara96,Hui99,McDonald99,Kujat01}.
This test is based on the fact that transverse (angular
separation) and radial (redshift separation) distances have a
different dependence on cosmological parameters, rendering a high
redshift spherical object in real space distorted in redshift
space. The degree of distortion increases with redshift and is
very sensitive to $\Lambda$ or, more generally, to dark energy. In
particular, Popowski \emph{et al.} \cite{Popowski98} (hereafter
PWRO) extended a calculation by Phillips \cite{Phillipps94} of the
geometrical distortion of the QSO correlation function. They
suggested a simple Monte Carlo experiment to see what constraints
should be expected from the 2dF QSO Redshift Survey (2QZ) and the
Sloan Digital Sky Survey (SDSS). However, they did not estimate
the probability density in the parameter space and, as a
consequence, they could not notice that the test is in fact very
sensitive to the difference $\Omega_{m0}-\Omega_{\Lambda 0}$.
Further, they did not take into account the effect of peculiar
velocities, although they discussed its role arguing that it would
not overwhelm the geometric signal.

Here we summarize the results which confirm the feasibility of
redshift distortion (geometric AP + peculiar velocity)
measurements to constrain cosmological parameters, by extending
the PWRO Monte Carlo experiments and obtaining confidence regions
in the ($\Omega _{m0},\Omega _{\Lambda 0} $) and ($\Omega_{m0},
w$) planes. We compare the expected constraints from the AP test,
when applied to the 2QZ survey, with those obtained by other
methods. We include a general dark energy component with equation
of state $P_{x}=w\,\rho_{x}$, with $w$ constant. Our analysis can
be generalized to dynamical scalar field cosmologies as well as to
any model with redshift dependent equation of state. Since most
quasars have redshift at around $z= 2$ we expect the test to be
useful in the determination of a possible redshift dependence of
the equation of state. We explicitly take into account the effect
of large-scale coherent peculiar velocities. Our calculations are
based on the measured 2QZ distribution function and we consider
best fit values for the amplitude and exponent of the correlation
function as obtained by Croom {\it et al.} \cite{Croom01}. In this
work, we only consider the 2QZ survey although the results can
easily be generalized to SDSS.

\section{Results and discussion}

\begin{figure}
\centering
    \includegraphics[height=14cm]{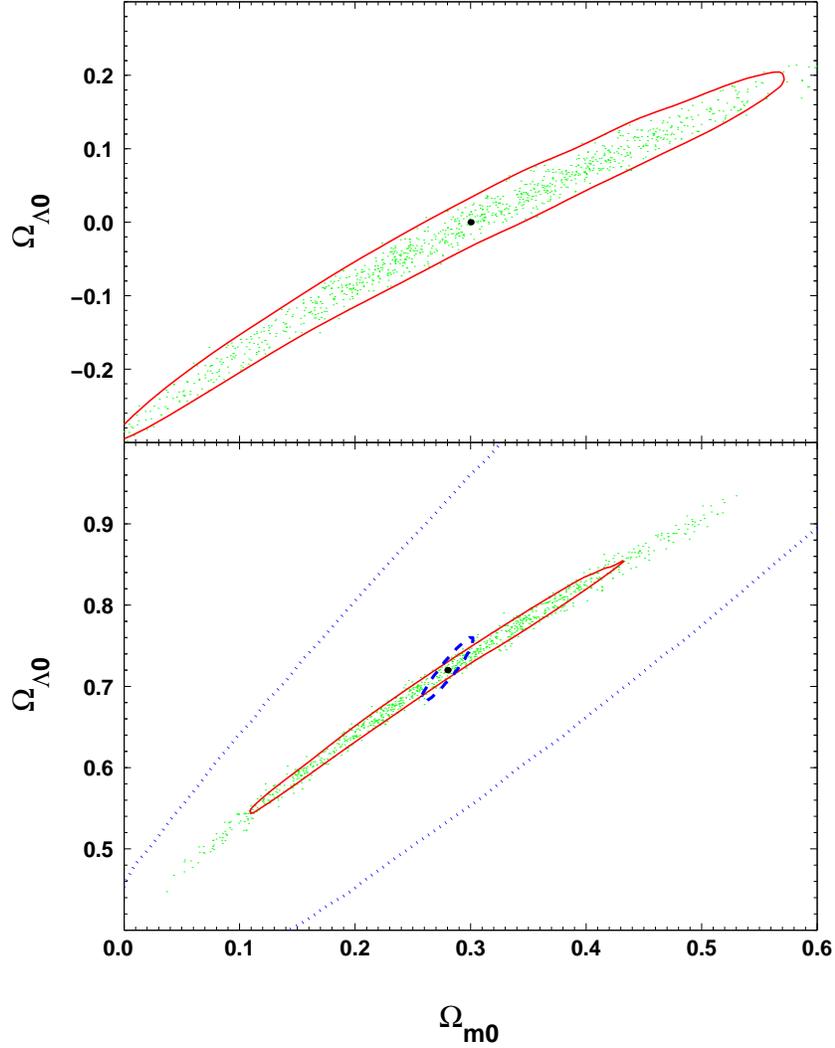}
\caption{Simulated models at fixed $w=-1$ and corresponding
predicted AP confidence contours (solid lines). In the top panel
we show the predicted $2\sigma$ likelihood contour assuming a
``true'' model ($\Omega_{m0}=0.3, \Omega_{\Lambda 0}=0)$. In the
bottom panel the predicted $1\sigma$ contour (dashed line) for one
year of SNAP data \protect\cite{Goliath01} is displayed, together
with the predicted $1\sigma$ AP contour. For both tests we
consider $\Omega_{m0}=0.28$ and $\Omega_{\Lambda 0}=0.72$; also
displayed is a $1\sigma$ confidence contour obtained by the
Supernova Cosmology Project (dotted lines;
\protect\cite{Perlmutter99}).}\label{fig:w-1}
\end{figure}

In Figure \ref{fig:w-1}, we show the predicted AP likelihood
contours in the ($\Omega_{m0},\Omega_{\Lambda 0}$)-plane for the
2QZ survey (solid lines), in the case $w=-1$, in a universe with
arbitrary spatial curvature. The scattered points represent
maximum likelihood best fit values for $\Omega_{m0}$ and
$\Omega_{\Lambda 0}$. The assumed ``true'' values are
($\Omega_{m0}=0.3,\Omega_{\Lambda 0}=0$) and ($0.28,0.72$), for
the top and bottom panels, respectively. In the top panel the
displayed curve corresponds to the predicted $2\sigma$ likelihood
contour. In the bottom panel the predicted $1\sigma$ contour
(dashed line) for one year of SNAP data \cite{Goliath01} is
displayed, together with the predicted $1\sigma$ AP contour. For
the SNAP contour, it is assumed that the intercept $\cal{M}$ is
exactly known. To have some ground of comparison with current SNe
Ia observations, in the same panel, we also plot (dotted lines)
the Supernova Cosmology Project \cite{Perlmutter99} $1\sigma$
contour (fit C). As expected, in both cases, the test recovers
nicely the ``true'' values. We stress out that the test is very
sensitive to the difference $\Omega_{m0}-\Omega_{\Lambda 0}$. From
the bottom panel we note that the sensitivity to this difference
is comparable to that expected from SNAP, of the order $\pm 0.01$.
Comparatively, however, the test has a larger uncertainty in the
determination of $\Omega_{m0}+\Omega_{\Lambda 0}$, of the order
$\pm 0.17$. The degeneracy in $\Omega_{m0}+\Omega_{\Lambda 0}$ may
be broken if we combine the estimated results for the AP test
with, for instance, those from CMB anisotropy measurements, whose
contour lines are orthogonal to those exhibited in the panels
\cite{Hu99}.

\begin{figure}
\centering \includegraphics[height= 14cm]{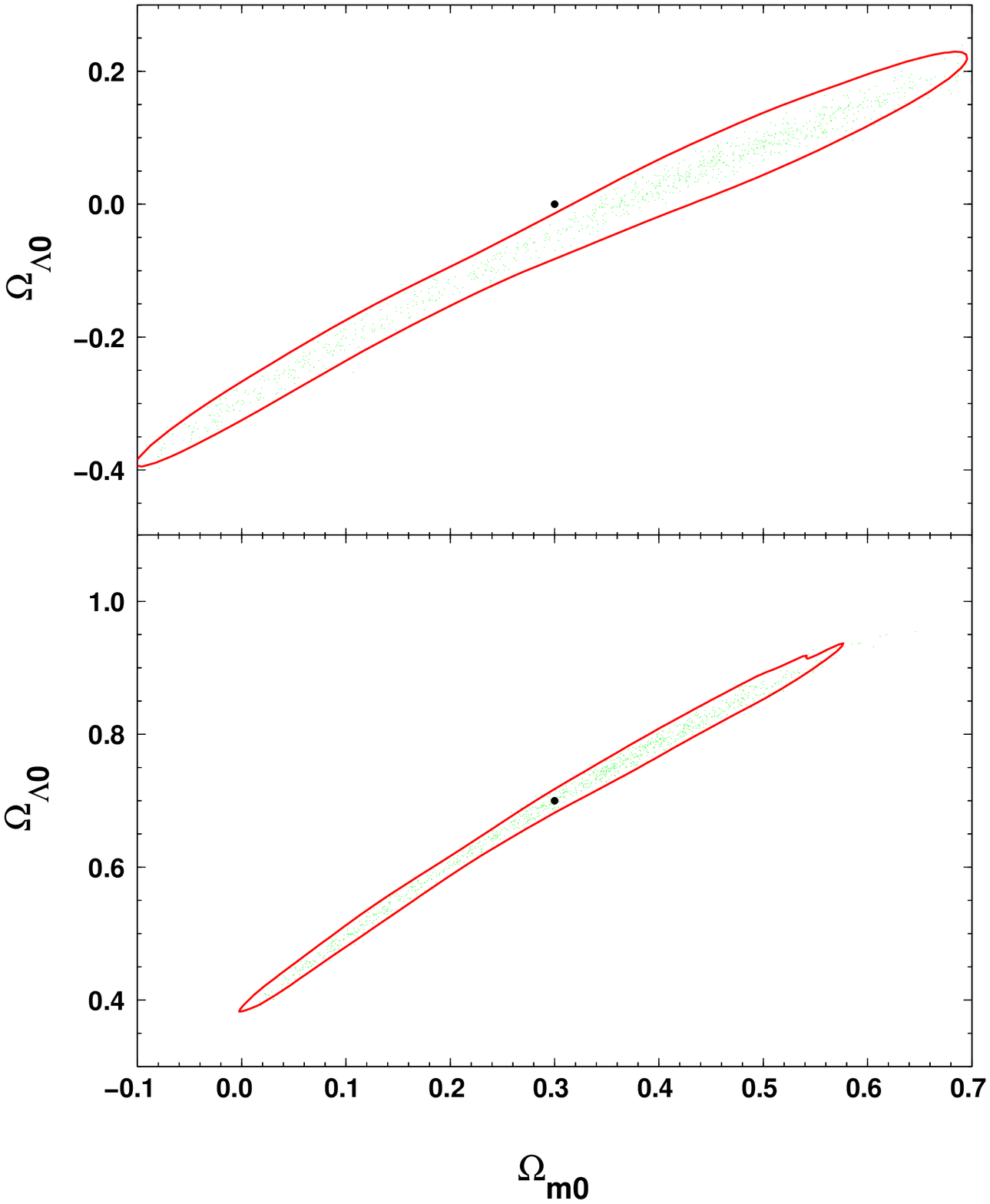}
\caption{Simulated models at fixed $w=-1$ and corresponding
$2\sigma$ predicted AP confidence contour; in both panels, the
``true'' model is indicated by a solid dot. Top panel: The
``true'' model, $(0.3, 0)$, takes into account the effect of
peculiar velocities, but the simulated ones do not. Notice that
the ``true'' model does not fall into the $2\sigma$ confidence
region. Bottom panel: The ``true'' model, $(0.3, 0.7)$, uses a
redshift dependent bias function with $b_0=1.45$ and $m=1.68$,
whereas the simulated ones use a constant bias equal to
2.46.}\label{fig:pvfavb}
\end{figure}

In order to estimate the consequences of neglecting the effect of
linear peculiar velocities, in the top panel of Figure
\ref{fig:pvfavb}, we included them in the calculation of the
$A_{i}$ values but neglected them in the computation of the
maximum likelihood; in this panel, we assume $\Omega _{m0}=0.3$
and $\Omega _{\Lambda 0}=0$ as ``true'' values. Notice that the
point with the ``true'' $\Omega _{m0}$ and $\Omega _{\Lambda 0}$
values is outside the $2\sigma $ contour. It is clear, therefore,
the necessity of taking this effect in consideration when
analyzing real data.

To illustrate that the AP test is in fact more sensitive to the
mean amplitude of the bias rather than to its exact redshift
dependence, we plot, in the bottom panel of Figure
\ref{fig:pvfavb}, the $2\sigma$ contour line, assuming as ``true''
values $\Omega_{m0}=0.3$ and $\Omega_{\Lambda 0}=0.7$. For this
panel, the ``true'' $A_i$ values were generated assuming
$b_0=1.45$ and $m=1.68$. However, for the simulations, we
considered a constant bias ($m=0$), such that
$b_{0,sim}:=\int_{z=z_{min}}^{z_{max}}F(z)b_{true}(z)dz=2.46$. We
remark that the contour is slightly enlarged, mainly in the
direction of the ``ellipsis'' major axis. However, the uncertainty
in $\Omega_{m0}-\Omega_{\Lambda 0}$ is practically unaltered,
confirming the strength of the test \cite{Yamamoto01}. We did the
same analysis assuming $\Omega_{m0}=1$ and $\Omega_{\Lambda 0}=0$
and obtained similar results.

\begin{figure}[htb]
\centering
\includegraphics[height=14cm]{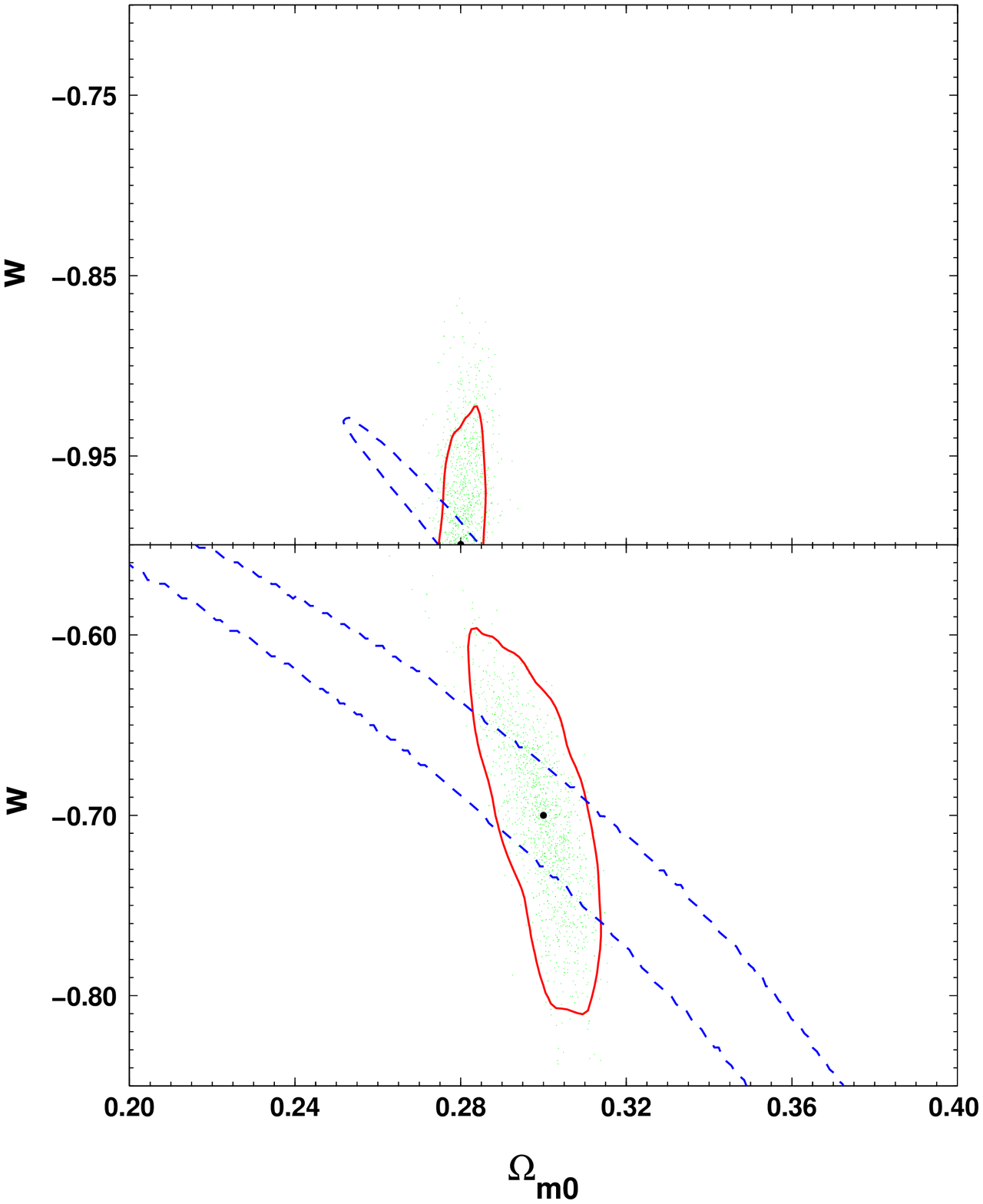}
\caption{Simulated flat models and corresponding
predicted AP confidence contours (solid lines). The top panel is
from a ``true'' model ($\Omega_{m0}=0.28$, $w=-1$), and displays
the predicted confidence contours for the AP test and the SNAP
mission (dashed line; \protect\cite{Goliath01}), both at $1\sigma$
level. The bottom panel is from a ``true'' model
($\Omega_{m0}=0.3$, $w=-0.7$), and displays the predicted
confidence contours for the AP test and the DEEP survey (dashed
line; \protect\cite{Newman00}), both at the $95\%$
level.}\label{fig:k0}
\end{figure}

In Figure \ref{fig:k0}, we show the predicted AP likelihood
contours in the $(\Omega_{m0},w)$-plane for the 2QZ survey (solid
lines) for flat models ($\Omega_{k0}=0$). The ``true'' values are
$(\Omega_{m0}=0.28,w=-1)$ and $(\Omega_{m0}=0.3,w=-0.7)$ for the
top and bottom panels, respectively. In the top panel, we show,
besides the AP contour, the predicted contour for one year of SNAP
data (dashed line; \cite{Goliath01}), both at $1\sigma$ level. For
the SNAP contour, the intercept $\cal{M}$ is assumed to be exactly
known. Notice that the contours are somewhat complementary and are
similar in strength. In the bottom panel, we compare the predicted
$95\%$ confidence contour of the AP test with the same confidence
contour for the number count test as expected from the DEEP
redshift survey (dashed line; \cite{Newman00}). Again the contours
are complementary, but the uncertainties on $\Omega_{m0}$ and $w$
for the AP test are quite smaller.

In summary, we have shown that the Alcock-Paczy\'nski test applied
to the 2dF quasar survey (2QZ) is a potent tool for measuring
cosmological parameters. We stress out that the test is especially
sensitive to $\Omega_{m0}-\Omega_{\Lambda 0}$. We have established
that the expected confidence contours are in general complementary
to those obtained by other methods and we again emphasize the
importance of combining them to constrain even more the parameter
space. We have also revealed that, for flat models, the estimated
constraints are similar in strength to those from SNAP with the
advantage that the 2QZ survey will soon be completed.

Of course our analysis can be improved in several aspects. For
instance, for the fiducial Einstein-de Sitter model, we have
assumed that $\gamma$ and $r_0$ do not depend on redshift. In
fact, observations \cite{Croom01} seem to support these
assumptions, but further investigations are necessary. Since the
test is very sensitive to $\Omega_{m0}-\Omega_{\Lambda 0}$, the
effect of small-scale peculiar velocities should also be
incorporated in future analyses in order to eliminate any
potential source of systematic bias. At present, the quasar
clustering bias is not completely well understood. Theoretical as
well as observational progress in its determination will certainly
improve the real capacity of the test. However, confirming
previous investigations \cite{Yamamoto01}, we have found that the
test is, in fact, more sensitive to the mean amplitude of the bias
rather than to its exact redshift dependence. A more extensive
detailed report of this work can be found in \cite{Calvao02}.

\Acknowledgements

We would like to thank J. Silk for calling attention to the
potential of the AP test and T. Kodama for suggestions regarding
numerical issues. We also thank the Brazilian research agencies
CNPq, FAPERJ and FUJB.


\begin{thebibliography}{99}

\bibitem{Riess99}  A. G. Riess \emph{et al.},
AJ {\bf 116}, 1009 (1999).

\bibitem{Perlmutter99}  S. Perlmutter \emph{et al.},
{ApJ} {\bf 517}, 565 (1999).

\bibitem{snap} Supernova Acceleration Probe: http://snap.lbl.gov/

\bibitem{deBernardis00}  P. de Bernardis \emph{et al.}, {Nature} {\bf 404}, 955
(2000).

\bibitem{Balbi00}  A. Balbi, \emph{et al.}, {ApJ},
{\bf 545}, L1 (2000).

\bibitem{Pryke01} C. Pryke \emph{et al.}, astro-ph/0104490 (2001).

\bibitem{map} Microwave Anisotropy Probe:
http://map.gsfc.nasa.gov/

\bibitem{planck} Planck Surveyor:
http://astro.estec.esa.nl/Planck/

\bibitem{2df} Two-Degree Field Survey: http://www.aao.gov.au/2df/

\bibitem{sdss} Sloan Digital Sky Survey: http://www.sdss.org/

\bibitem{Ratra88}  B. Ratra, and P. J. E. Peebles {Phys. Rev. D}, {\bf 37},
3406(1988).

\bibitem{Frieman95}  J. A. Frieman, C. T. Hill, A. Stebbins, and I. Waga, {Phys. Rev. Lett.},
{\bf 75}, 2077 (1995).

\bibitem{Caldwell98}  R. R. Caldwell, R. Dave, and P. J. Steinhardt, {Phys. Rev. Lett.}, {\bf 80},
1582 (1998).

\bibitem{Ferreira98}  P. Ferreira, and M. Joyce, {Phys. Rev. D}, {\bf 58}, 023503 (1998).

\bibitem{Alcock79}  C. Alcock, and B. Paczy\'nski, {Nature}, {\bf 281}, 358 (1979) [AP].

\bibitem{Ryden95} B. S. Ryden, {ApJ}, {\bf 452}, 25 (1995).

\bibitem{Ballinger96}   W. E. Ballinger, J. A. Peacock, and A. F. Heavens, {MNRAS},
{\bf 282}, 877 (1996).

\bibitem{Matsubara96}  T. Matsubara, and Y. Suto, ApJ, {\bf 470}, L1 (1996).

\bibitem{Hui99} L. Hui, A. Stebbins, and S. Burles, {ApJ}, {\bf 511}, L5
(1999).

\bibitem{McDonald99} P. McDonald, and J. Miralda-Escud\'e, {ApJ}, {\bf 518}, 24
(1999).

\bibitem{Kujat01} J. Kujat, A. M. Linn, R. J. Scherrer, and D. H.
Weinberg, astro-ph/0112221 (2001)

\bibitem{Popowski98}  P. A. Popowski, D. H. Weinberg, B. S. Ryden, and
P. S. Osmer, {ApJ}, {\bf 498}, 11 (1998) [PWRO].

\bibitem{Phillipps94}  S. Phillipps,
{MNRAS}, {\bf 269}, 1077 (1994).

\bibitem{Croom01}  S. M. Croom {\it et al}., MNRAS, {\bf 325}, 483 (2001).

\bibitem{Goliath01} M. Goliath, R. Amanullah, P. Astier, A. Goobar, and
R. Pain, Astron. Astrophys., {\bf 380}, 6 (2001).

\bibitem{Hu99} W. Hu, D. J. Eisenstein, M. Tegmark, and M. White,
{Phys. Rev. D}, {\bf 59}, 023512 (1999).

\bibitem{Yamamoto01} K. Yamamoto, and H. Nishioka, ApJ, {\bf 549}, L15 (2001).

\bibitem{Newman00} J. F. Newman, and M. Davis, ApJ, {\bf 534}, L11 (2000).

\bibitem{Calvao02} M. O. Calv\~ao, J. R. T. de Mello Neto, and I.
Waga, Phys. Rev. Lett. (in press); astro-ph/0107029


\end{thebibliography}
\end{document}